\documentclass{article}

\usepackage[utf8]{inputenc}
\usepackage[T1]{fontenc}
\usepackage{lmodern}
\input{glyphtounicode}\pdfgentounicode=1
\usepackage{amsmath,amssymb}
\usepackage{graphicx}
\usepackage{booktabs}
\usepackage{longtable}
\usepackage[colorlinks=true,linkcolor=blue!50!black,citecolor=blue!50!black,urlcolor=blue!50!black]{hyperref}
\usepackage[margin=2cm]{geometry}
\usepackage{natbib}
\usepackage{xcolor}
\usepackage{placeins}
\usepackage{float}
\usepackage{authblk}

\newcommand{\resultsdir}{figures}

\title{CTseg: A Tool for Brain CT Segmentation,\\Spatial Normalisation, and Volumetrics}
\author{Mikael Brudfors}
\affil{Wellcome Centre for Human Neuroimaging, University College London}
\date{}

\begin{document}
\maketitle

\begin{abstract}
This paper presents and validates CTseg, a freely available software for brain CT
segmentation, spatial normalisation, and volumetrics. CTseg builds on the
Multi-Brain generative modelling framework, providing a CT-specific pipeline that produces tissue
maps, deformation fields, and brain volume estimates in the same format
as SPM's unified segmentation, thereby extending SPM's
established analysis chain from MRI to CT. CTseg is designed for routine
hospital CT scans without requiring preprocessing or resampling in
deployment. Although CTseg has been adopted in clinical research spanning, among other things,
stroke, dementia, and brain morphometry, a systematic validation against
an independent reference standard has been lacking. Using paired MR/CT
head scans, we
evaluate CTseg across four dimensions: segmentation accuracy against an MRI-derived silver
standard; spatial normalisation consistency through group-average
sharpness and voxelwise coefficient of variation; brain volume agreement via intraclass correlation and Bland-Altman analysis; and
downstream sex classification performance from normalised tissue maps.
As a baseline, we apply SPM's MRI-based unified
segmentation directly to the CT images. CTseg significantly outperformed this baseline for segmentation and normalisation, showed stronger TBV agreement, and achieved comparable TIV agreement. CTseg is freely available at
\url{https://github.com/WCHN/CTseg}, and all experiment code is included
in the repository for full reproducibility.
\end{abstract}

\section{Introduction}
\label{sec:intro}

Computed tomography (CT) is the most frequently acquired neuroimaging
modality in clinical practice. It is the first-line investigation for
stroke, traumatic brain injury, and intracranial haemorrhage, and is
routinely obtained in emergency, neurosurgical, and oncological settings.
Across most healthcare systems, head CT exams substantially outnumber
brain MRI exams~\citep{smithbindman2019trends}, yet the dominant computational neuroimaging analysis tools have been developed for MRI.

Being able to automatically segment brain tissue and spatially normalise CT
scans to a standard coordinate
system~\citep{friston1995spatial} would unlock significant clinical and
research value. It would enable population-level analyses such as
voxel-based morphometry~\citep{ashburner2000vbm} on the large volumes of
CT data that already exist in hospital archives, support quantitative
assessment of brain atrophy from routine clinical scans, facilitate lesion
mapping in standard space for stroke and trauma studies, and provide
automated brain volume estimates (total brain volume, TBV, and total
intracranial volume, TIV) that are recognised as clinical biomarkers. These capabilities are well established for MRI,
where software suites such as SPM~\citep{tierney2025spm},
FreeSurfer~\citep{fischl2012freesurfer}, and
FSL~\citep{jenkinson2012fsl} provide mature, validated pipelines. However,
applying these MRI-trained tools to CT data is problematic because the
tissue contrast in CT is fundamentally different: the Hounsfield unit scale
reflects X-ray attenuation, yielding substantially poorer differentiation
between grey matter (GM), white matter (WM), and cerebrospinal fluid (CSF)
than MRI.

In this paper, we present and validate
CTseg\footnote{Not to be confused with the unrelated ``CTSeg'' tool by
\citep{adduru2020ctseg}, which uses a different
approach.}, a CT-specific tool
that addresses these limitations. CTseg can be used within an
SPM-compatible framework, without supervised labels or GPU dependence. Although CTseg has already been adopted in
clinical research across diverse populations, a
systematic validation against an
independent reference standard has not yet been performed. We validate CTseg using 59 paired MR/CT head scans from the SynthRAD2025
challenge~\citep{thummerer2025synthrad}. The paired design enables us to
use SPM's MRI segmentation~\citep{ashburner2005unified} as a silver
standard. As a baseline, we also apply SPM's unified segmentation directly
to the CT images, providing a controlled head-to-head comparison that
isolates the benefit of CT-specific modelling. SPM is the appropriate
primary comparator because both CTseg and SPM are unsupervised, model-based
methods within the same framework. Deep learning methods for CT
segmentation typically require labels and generally address a narrower
task: they produce segmentations but not the deformation fields or
volumetric outputs needed for spatial normalisation, voxel-based
morphometry, or brain volume estimation. A comparison with such
methods is therefore outside the scope of this validation but would be
valuable future work.
\section{Related Work}
\label{sec:related}

Automated analysis of brain CT has a shorter history than its MRI
counterpart, largely because the poor soft-tissue contrast in CT was long
considered insufficient for tissue-level segmentation. Early work focused
primarily on spatial normalisation. \citet{rorden2012age} developed
stroke-aged CT and MRI atlases and packaged them in the SPM Clinical
Toolbox, enabling affine and low-dimensional nonlinear normalisation of
clinical CT scans for lesion mapping. \citet{muschelli2020template} later
produced a higher-resolution, unbiased CT atlas using an iterative
registration procedure, together with coarse ventricle and whole-brain
labels. Both approaches are atlas and normalisation resources rather
than subject-level CT tissue-segmentation methods.

Tissue segmentation from CT has been addressed by a smaller body of work.
\citet{gupta2010segmentation} proposed an early approach for automatic
GM, WM, and CSF segmentation from unenhanced CT using adaptive
thresholding and anatomical knowledge, and
\citet{kemmling2012decomposing} introduced a probabilistic Hounsfield-unit
decomposition method using MRI-derived tissue maps and a custom CT
atlas. \citet{manniesing2017segmentation} demonstrated that GM and WM
can be segmented from 4D contrast-enhanced CT by exploiting temporal
averaging across a perfusion series, followed by voxel feature extraction,
SVM classification, and atlas-guided geodesic active contours. While
effective for contrast-enhanced acquisitions, this method is not applicable
to the far more common single-phase non-contrast CT.

More recently, deep learning methods have been applied to the problem.
\citet{srikrishna2021deeplearning} trained a U-Net on paired CT/MRI data
from the Gothenburg H70 Birth Cohort, using MRI-derived FreeSurfer labels
as supervision, and achieved Dice scores of 0.79 (GM), 0.82 (WM), and
0.75 (CSF) on held-out CT images. \citet{cai2020fully} used a
deep-learning model to segment 11 intracranial structures from
non-contrast head CT, targeting radiological reporting rather than
tissue-class analysis. \citet{gerken2023deeplearning} trained a 2D U-Net
for brain parenchyma and ventricular segmentation on CT, demonstrating
robustness to pathology including haemorrhage and tumour.
\citet{son2024automated} used perceptual-loss U-Nets trained on 199 paired
CT/MRI scans for GM, WM, and CSF segmentation. Most recently,
\citet{huisman2024synthseg} validated SynthSeg, a domain-randomised
segmentation tool originally developed for MRI, on CT using 260 paired
CT/MRI radiotherapy scans from five centres. These supervised and
domain-randomised methods are segmentation-centric and do not themselves
provide nonlinear deformation fields or an SPM-compatible end-to-end
morphometry pipeline. Separately, \citet{fielden2022volumetrics} compared
global CT-versus-MRI brain volumetrics using SPM and FSL in 69 paired
scans, providing an independent reference point for CT-based volume
estimation.

CTseg occupies a distinct position in this landscape: it is a generative
model-based method that jointly estimates tissue segmentation and nonlinear
spatial normalisation from a single non-contrast CT scan, without requiring
labelled training data, contrast enhancement, or preprocessing. Its outputs
are compatible with the standard SPM analysis pipeline, bridging the gap
between CT acquisition and the quantitative analysis tools that are well
established for MRI.

\subsection{Prior Applications of CTseg}
\label{sec:adoption}

Although CTseg was initially released without a formal validation study, it has
since been adopted in a range of clinical research settings.
Table~\ref{tab:adoption} summarises 13 published studies, collectively
encompassing over 5{,}000 subjects, that have used CTseg or CTseg-derived
outputs. Applications span stroke outcome prediction and perfusion
mapping, brain morphometry in Down syndrome, COVID-19 neuroimaging,
cardiac arrest prognosis, radiomics, PET imaging, bone density estimation,
mortality prediction, and dementia screening. These studies demonstrate
the practical utility of CTseg across diverse clinical populations, but
they do not substitute for direct validation against an independent
reference standard -- the present study provides such a validation.
 
 \begin{longtable}{p{2.8cm}p{2.2cm}p{2.4cm}p{0.8cm}p{6.5cm}}
 	\caption{Published external applications of CTseg, as of April 2026.}
 	\label{tab:adoption} \\
 	\toprule
 	Study & Domain & CTseg Use & $N$ & Key Finding \\
 	\midrule
 	\endfirsthead
 	\toprule
 	Study & Domain & CTseg Use & $N$ & Key Finding \\
 	\midrule
 	\endhead
 	\midrule
 	\multicolumn{5}{r}{\small\emph{Continued on next page}} \\
 	\endfoot
 	\bottomrule
 	\endlastfoot
 	\citet{liu2021atrophy}
 	& Basilar artery occlusion
 	& BPV, CSF volume, brain atrophy index
 	& 231
 	& Severe brain atrophy (estimated from CT via CTseg) independently
 	predicted poor functional outcome after endovascular treatment in the
 	BASILAR registry. \\[4pt]
 	\citet{hoving2024impact}
 	& Ischaemic stroke
 	& ICV, TBV from baseline NCCT
 	& 200
 	& Adjusting CT perfusion core volume by CTseg-derived ICV or TBV did not
 	improve prognostic value beyond absolute core volume in the MR CLEAN
 	Registry. \\[4pt]
 	\citet{luijten2026prediction}
 	& Ischaemic stroke (EVT)
 	& Brain parenchymal fraction
 	& 1,391
 	& CT-based prediction model incorporating CTseg-derived brain atrophy
 	improved identification of patients benefiting from thrombectomy across
 	7 randomised trials. \\[4pt]
 	\citet{sanchezmoreno2024vbm}
 	& Down syndrome
 	& VBM, tissue volumes, spatial normalisation
 	& 98
 	& First CT-based VBM study in Down syndrome; CTseg-derived GM volumes
 	correlated with dementia severity and plasma AD biomarkers (p-tau181,
 	NfL) in hippocampal and cingulate regions. \\[4pt]
 	\citet{sanchezmoreno2025ctmri}
 	& Down syndrome
 	& CT vs MRI morphometry comparison
 	& 23
 	& 92.3\% agreement between CTseg (CT) and SPM DARTEL (MRI) morphometry;
 	similar patterns of regional volume loss associated with cognitive
 	decline. \\[4pt]
 	\citet{urbanos2025radiomics}
 	& Subarachnoid haemorrhage
 	& GM/WM segmentation for radiomics feature extraction
 	& 403
 	& Radiomics features extracted from CTseg tissue segmentations predicted
 	6-month mortality, functional outcome (GOS), vasospasm, and
 	hydrocephalus in SAH patients. \\[4pt]
 	\citet{kalc2024bone}
 	& Bone mineral density
 	& CT skull segmentation for validation
 	& 114
 	& CTseg skull segmentation of OASIS-3 CT scans used as ground truth
 	to validate MRI-derived proxy measures of bone mineral density and
 	subcutaneous adiposity. \\[4pt]
 	\citet{tsui2023mortality}
 	& Acute hospital mortality
 	& TBV, TIV from routine NCCT
 	& 804
 	& CT-derived brain volumes integrated into a multidimensional clinical
 	model predicted mortality in acutely hospitalised older patients,
 	demonstrating prognostic value of routinely acquired CT data. \\[4pt]
 	\citet{gramespacher2024cardiac}
 	& Cardiac arrest outcome
 	& GM segmentation from NCCT
 	& 132
 	& CTseg GM segmentation of cerebral CT used as input to a supervised
 	ML classifier predicting neurological outcome after out-of-hospital
 	cardiac arrest. \\[4pt]
 	\citet{tangwiriyasakul2026stroke}
 	& Ischaemic stroke perfusion
 	& Spatial normalisation, GM/WM atlases
 	& 1,393
 	& CTseg used to co-register and nonlinearly normalise routine NCCT and
 	CTA to MNI space; CTseg GM/WM atlases used to segment derived
 	perfusion-deficit substrates in a deep generative model. \\[4pt]
 	\citet{duan2021covid}
 	& COVID-19 neuroimaging
 	& VBM, tissue volumes, spatial normalisation
 	& 120
 	& CT images normalised to MNI and segmented into six tissue classes
 	using CTseg; VBM revealed frontal-temporal GM volume reductions
 	associated with clinical severity in older COVID-19 patients. \\[4pt]
 	\citet{tonietto2024tauPET}
 	& Alzheimer's tau PET imaging
 	& Skull/meninges segmentation from CT
 	& 18
 	& CTseg skull probability maps from PET attenuation-correction CT used
 	to define skull/meninges ROI for isolating off-target tau tracer
 	binding in a head-to-head comparison of [{$^{18}$F}]PI-2620 and
 	[{$^{18}$F}]RO948. \\[4pt]
 	\citet{yelanchezian2025dementia}
 	& Dementia screening
 	& TBV, hippocampal volume
 	& 168
 	& CTseg-derived TBV and hippocampal volume from routine clinical CT
 	differentiated dementia from non-dementia (combined AUC\,=\,0.74) in a
 	New Zealand memory service. \\
 	\midrule
 	\multicolumn{3}{r}{\textbf{Total}} & \textbf{5{,}095} & \\
 \end{longtable}
 
\section{Methods}
\label{sec:ctseg}

\subsection{Overview}

CTseg is a CT-specific pipeline for simultaneous tissue segmentation and
nonlinear spatial normalisation. It builds on the Multi-Brain
generative modelling
framework~\citep{brudfors2020flexible}\footnote{Source code:
	\url{https://github.com/WTCN-computational-anatomy-group/mb}}, which
extends SPM's unified segmentation
approach~\citep{ashburner2005unified} with improved affine and nonlinear
registration and jointly learned atlas and intensity priors. The CTseg model
produces six tissue class segmentations (GM, WM, CSF, Bone, Soft tissue,
and Background) together with nonlinear deformation fields that map each
scan to an atlas space. In addition, CTseg provides
skull-stripping and estimates of TBV and TIV.

CTseg works directly on clinical CT scans without preprocessing. It
handles arbitrary voxel sizes (including the variable slice thickness
common in clinical CT), arbitrary fields of view, and does not assume any
particular orientation or prior alignment. Because CTseg produces outputs
in the same format as SPM's unified segmentation (native and
atlas-space tissue maps, deformation fields, and modulated images), it
serves as a drop-in replacement that extends SPM's downstream analysis
tools, including voxel-based
morphometry~\citep{ashburner2000vbm}, atlas-based region-of-interest
analysis~\citep{tzourio2002aal}, and voxel-based lesion-symptom
mapping~\citep{bates2003vlsm}, from MRI to CT data.

\subsection{Training Data}

The CTseg atlas and intensity priors were learned jointly from both CT
and MRI data. The training set comprised four publicly available
datasets:

\begin{itemize}
	\item \textbf{CQ500}~\citep{chilamkurthy2018cq500}: 491 non-contrast
	head CT scans, of which the 222 scans reported as normal by all three
	radiologist readers (no haemorrhage, fracture, mass effect, or
	midline shift) were used for atlas construction. In-plane resolution
	$\sim$0.5\,mm, variable slice thickness (0.625-5.0\,mm).
	\item \textbf{IXI}\footnote{\url{https://brain-development.org/ixi-dataset}}: 577 multi-sequence MR images
	(T1w, T2w, PDw) from healthy subjects, acquired at three London
	hospitals, $\sim$1\,mm isotropic.
	\item \textbf{MICCAI 2012 Multi-Atlas
		Labeling}~\citep{landman2012miccai}: 35 T1-weighted MRI scans (30
	unique subjects, 5 rescanned) with expert manual segmentations of 133
	brain regions, 1\,mm isotropic. The manual labels were combined into GM
	and WM tissue classes and used to supervise the atlas learning.
	\item \textbf{MRBrainS18}~\citep{kuijf2024mrbrains18}: 7 training
	subjects with multi-sequence MRI (T1w, T1-IR, T2-FLAIR) and manual
	segmentations of 8 brain structures, from subjects aged over~50,
	including patients with pathologies. The manual labels were combined
	into GM (cortical grey matter + basal ganglia), WM (white matter +
	white matter lesions), and CSF (cerebrospinal fluid + ventricles) to
	supervise the atlas learning. Pathological tissue (e.g.\ infarctions)
	was masked out during training.
\end{itemize}

The inclusion of both modalities is a key design choice: MRI provides
superior soft-tissue contrast for learning tissue boundaries, while CT
contributes bone structures and ensures the intensity priors capture
Hounsfield-unit profiles. Where manual labels were available
(MICCAI~2012, MRBrainS18), they guided the merging of learned Gaussian
components into tissue classes; the remaining datasets (CQ500, IXI)
contributed unlabelled images, with tissue classes inferred entirely from
the generative model. All training images were rigidly aligned to MNI space prior to model
fitting, which speeds up convergence of learning the Multi-Brain model.

\subsection{Atlas Learning}

\begin{figure}[H]
	\centering
	\IfFileExists{\resultsdir/fig_ctseg_intensity_priors.pdf}{%
		\includegraphics[width=0.8\columnwidth]{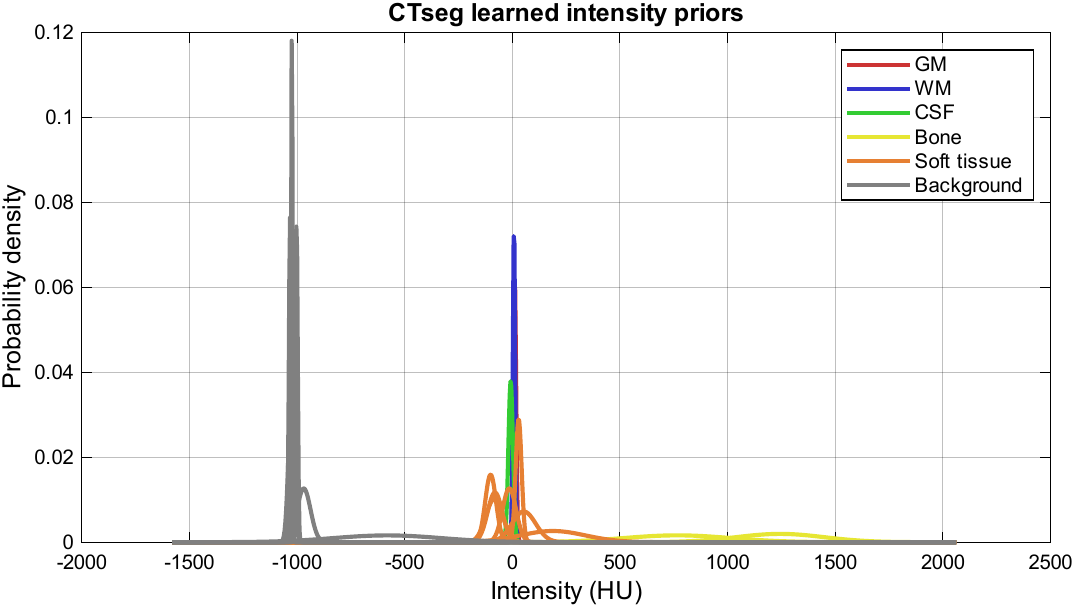}%
	}{\fbox{\parbox{\columnwidth}{\centering\vspace{2cm}fig\_ctseg\_intensity\_priors.pdf\vspace{2cm}}}}
	\caption{Learned $K = 18$ GMM intensity priors for CT data.
		Each colour represents one of the six tissue classes. Individual
		Gaussian components within each class are shown as separate curves of
		the same colour. For example, the background class (grey) comprises
		six Gaussian components, reflecting distinct intensity sub-populations
		(e.g.\ air, scanner bed).}
	\label{fig:priors}
\end{figure}

The Multi-Brain framework jointly estimates a tissue atlas,
Gaussian mixture model (GMM) intensity priors, and nonlinear deformation
fields within a single generative model. The model was configured with 1mm atlas voxel size and 
$K = 18$ Gaussian components, and fit to the pooled training data using
default hyper-parameters for the diffeomorphic registration
regularisation, bias field correction, etc. After convergence, the 18 components
were merged by visual inspection into six tissue classes. Each class is
represented by a single spatial prior (one channel of the atlas) but may
be modelled by multiple Gaussian components in the intensity domain. For
example, the background class is captured by a single spatial prior map
but six Gaussians, reflecting the distinct intensity profiles of air,
surrounding tissue, and scanner-specific artefacts
(Figure~\ref{fig:priors}). This GMM-per-class structure allows each
tissue class to capture multimodal intensity distributions without
requiring additional spatial prior channels. Full details of the model
specification, optimisation procedure, and convergence criteria are
provided in~\citep{brudfors2020phd}. Learning the CTseg model took 24 hours on a 24-core workstation with 128\,GB of RAM.

\subsubsection{Original Groupwise Optimal Atlas}

\begin{figure}[H]
	\centering
	\IfFileExists{\resultsdir/fig_ctseg_template_original.pdf}{%
		\includegraphics[width=0.9\textwidth]{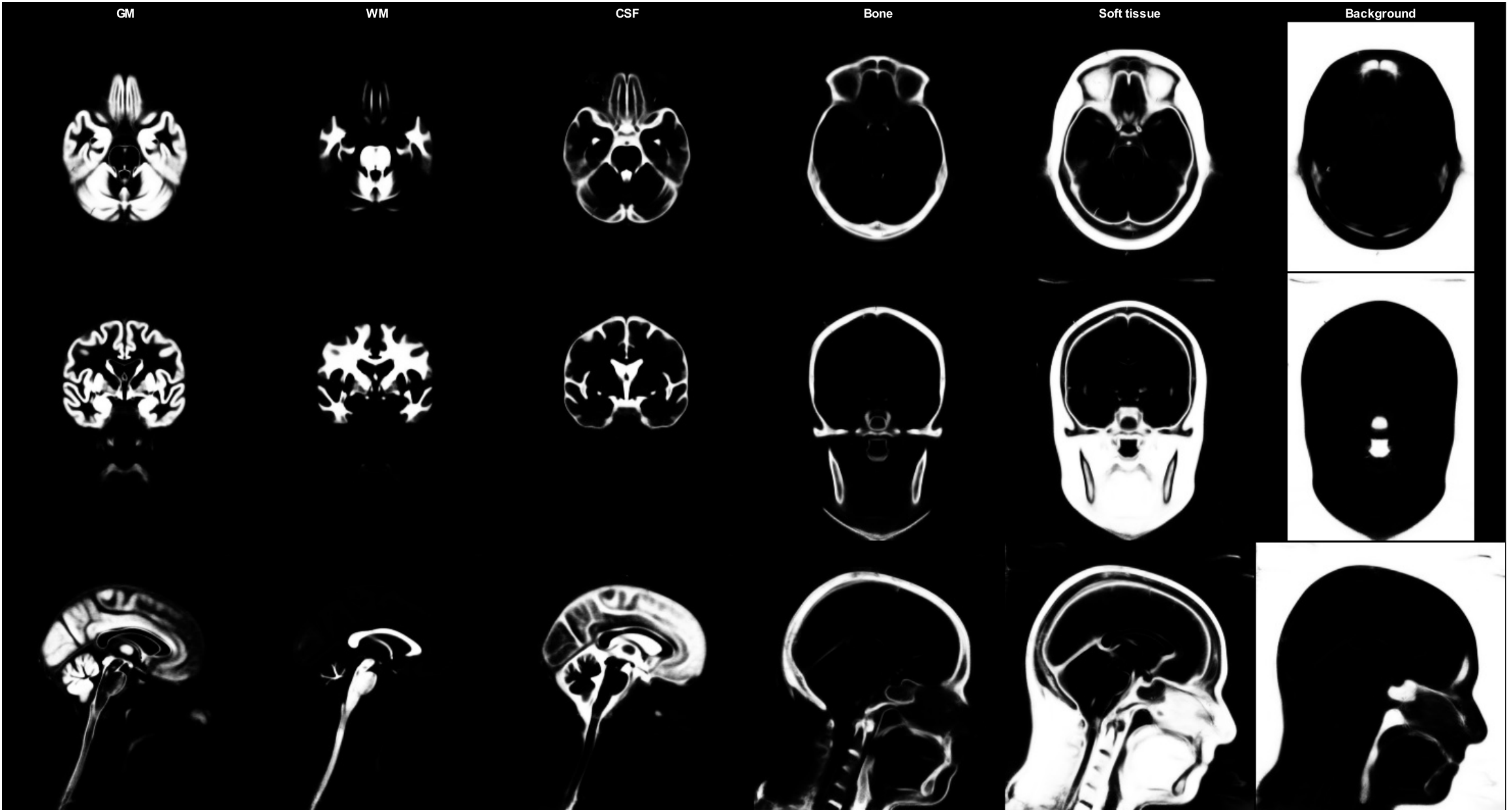}%
	}{\fbox{\parbox{\textwidth}{\centering\vspace{2cm}fig\_ctseg\_template\_original.pdf\vspace{2cm}}}}
	\caption{Original groupwise CTseg tissue atlas
		(\texttt{mu\_CTseg.nii}), learned in a data-driven coordinate system.
		Note the extended caudal field of view covering the upper cervical spine.
		Columns: tissue classes (GM, WM, CSF, Bone, Soft tissue, Background). Rows: axial, coronal, and sagittal views. Brighter voxels indicate
		higher prior probability.}
	\label{fig:template_original}
\end{figure}

The resulting atlas \texttt{mu\_CTseg.nii} was learned in a data-driven coordinate system that
is optimal for the training population (Figure~\ref{fig:template_original}). A notable
feature is its extended caudal field of view, which includes the upper
cervical spine. This benefits processing of clinical CT scans that
frequently extend below the skull base into the neck, as the registration
model can anchor to inferior anatomy and reduce misregistration risk when
the brain occupies only a portion of the image volume.

\subsubsection{SPM-aligned Atlases}
\label{sec:spm_atlases}

The CTseg atlas described above was learned in a groupwise-optimal
space that does not coincide with the space of the SPM atlas. This is a consequence of the
Multi-Brain framework, which learns the atlas jointly with the
registration parameters and is therefore not constrained to any
predefined coordinate system. While this groupwise space is optimal
for the training population, it means CTseg's deformation fields do
not map directly to, e.g., SPM atlas space. To address this, we nonlinearly align the CTseg atlas to SPM's tissue
probability space (\texttt{TPM.nii}). Rather than routing the alignment
through an intermediate MRI reference image, we register
\texttt{mu\_CTseg.nii} directly to \texttt{TPM.nii} with Multi-Brain,
using the tissue-probability channels of both atlases as categorical
inputs. Because the source and target are already tissue-probability
fields, no intensity model is needed: the registration matches
posteriors to posteriors, which is both simpler and more faithful than
aligning the atlas to a single-subject MR intensity average. The
resulting deformation is then applied to warp the CTseg template into
SPM space at the requested voxel size.

Two SPM-aligned atlases are distributed with CTseg:
\begin{itemize}
	\item \texttt{mu\_CTseg\_spm15.nii}: 1.5\,mm isotropic resolution,
	matching the voxel size of \texttt{TPM.nii}. This is the default
	atlas and the version used throughout this paper.
	\item \texttt{mu\_CTseg\_spm10.nii}: 1.0\,mm isotropic resolution,
	for applications requiring higher spatial fidelity at the cost of
	longer processing time.
\end{itemize}
Because the SPM-aligned atlases live directly in SPM space,
\texttt{wc*}/\texttt{mwc*} outputs can be analysed with standard SPM
tooling without an additional warp through \texttt{spm\_CTseg\_warp}.


\section{Validation Study}
\label{sec:methods}

\subsection{Data}
\label{sec:data}

Validating a CT segmentation tool against an MRI-derived reference requires
a dataset of spatially aligned, same-subject MR and CT brain scans. Such
datasets are rare in the public domain, as MR and CT are typically
acquired for different clinical indications and CT involves ionising
radiation that is difficult to justify for research alone. The few existing
paired datasets come from radiotherapy planning, where both modalities are
acquired as part of routine care. We therefore used the SynthRAD2025 Grand
Challenge dataset~\citep{thummerer2025synthrad}, which is, to our knowledge, among the largest publicly available paired MR-CT datasets.
The SynthRAD2025 dataset comprises 2{,}362 paired MRI-CT and CBCT-CT
cases from head-and-neck, thoracic, and abdominal cancer patients, acquired
at five European university medical centres (UMC Groningen, UMC Utrecht,
Radboud UMC, LMU University Hospital Munich, and University Hospital of
Cologne) using diverse scanners and acquisition protocols. From this
dataset, we used the Task~1 (Head \& Neck) MRI-CT training subset. The
full training set comprises 221 subjects from three centres (A, C, and~D);
however, centre~D data (65 subjects) is distributed under a restricted
licence that limits its use to the duration of the challenge, so we used
the 156 freely available subjects from centres A ($n$\,=\,91) and
C ($n$\,=\,65), each with a paired MR and CT scan.

The challenge organisers pre-processed all images prior to
release~\citep{thummerer2025synthrad}: (1)~MR images were rigidly
registered to the corresponding planning CT using Elastix; (2)~facial
structures were removed (defaced) for anonymisation; (3)~images were
resampled to $1 \times 1 \times 3$\,mm voxel spacing; and (4)~images were
cropped to the patient outline. No bias field correction or intensity
normalisation was applied to the MR images. Images were provided in
MetaImage format (.mha) and stored as INT16. Notably, deformable
registration of the CT to the MR, to correct for anatomical differences
between the two scan sessions, was only performed by the challenge
organisers for the validation and test sets, not for the training data.
We therefore replicated this step ourselves (described in
Section~\ref{sec:preprocessing}).

\begin{figure}[t]
    \centering
    \IfFileExists{\resultsdir/fig_example_data.pdf}{%
        \includegraphics[width=0.8\columnwidth]{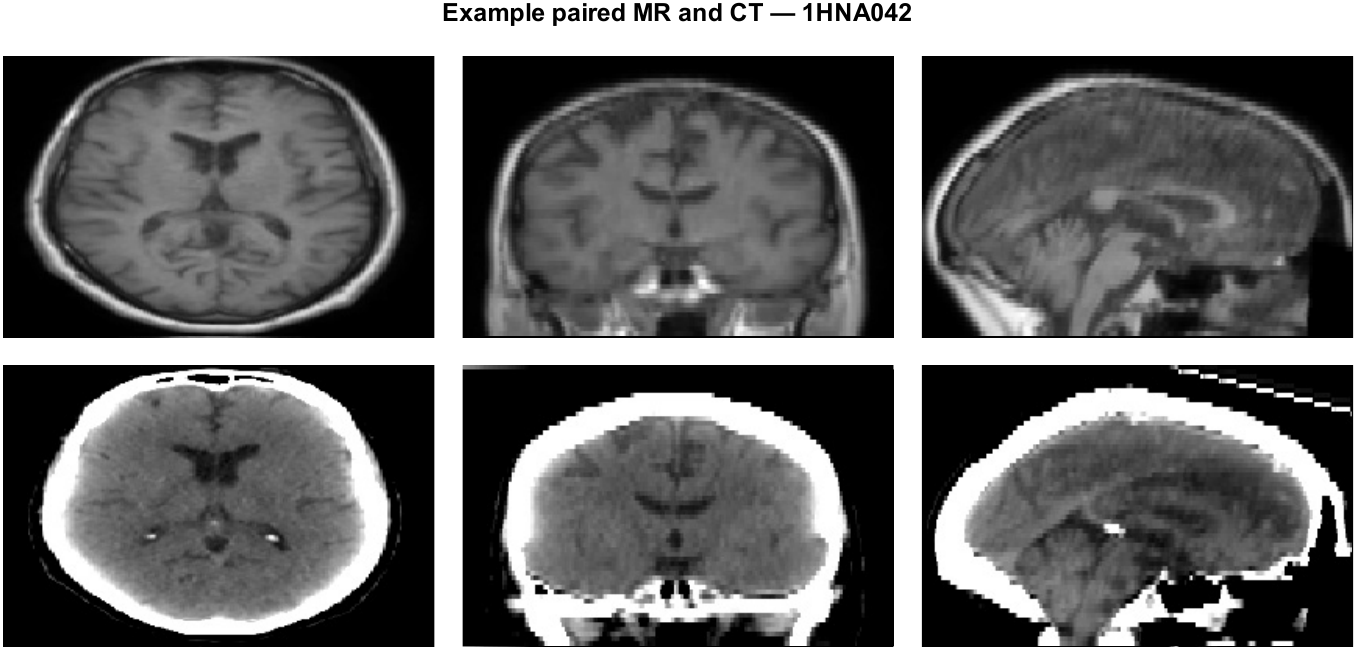}%
    }{\fbox{\parbox{\columnwidth}{\centering\vspace{2cm}fig\_example\_data.pdf\vspace{2cm}}}}
    \caption{Example paired MR (top) and CT (bottom) images from the
    SynthRAD2025 dataset (Task~1 Head \& Neck, MRI-CT training subset), shown in axial, coronal, and sagittal views.}
    \label{fig:example_data}
\end{figure}

The accompanying metadata provides patient demographics and acquisition
parameters. Of the 156 subjects, 64 were male, 27 female, and 65 had sex
unlisted; age was available for 90 subjects (mean 64 years, range
29-87). CT scans were acquired on scanners from two manufacturers
(Philips, $n$\,=\,118; Siemens, $n$\,=\,38) across four scanner models, at
120\,kVp, with slice thicknesses of 1-3\,mm. Figure~\ref{fig:example_data} shows example MR
and CT views from a representative subject.

\subsection{Preprocessing}
\label{sec:preprocessing}

All images were preprocessed using the \texttt{spm-hospital-preproc}
toolbox\footnote{\url{https://github.com/WTCN-computational-anatomy-group/spm-hospital-preproc}}.
Preprocessing consisted of four steps: (1)~rigid-body alignment of each
image to MNI space; (2)~co-registration of the MR and CT images within
each subject pair using normalised mutual information, to ensure precise
spatial correspondence; (3)~cropping of both images to the SPM12 atlas
field of view ($181 \times 217 \times 60$ voxels at the original
$1 \times 1 \times 3$\,mm voxel size); and (4)~deformable registration
of the CT to the MR to correct for residual anatomical differences
between the two scan sessions (e.g.\ from weight change or organ motion),
replicating the procedure used by the challenge organisers for the
validation and test sets~\citep{thummerer2025synthrad}. The deformable
registration used the Elastix framework~\citep{klein2010elastix} via
SimpleITK with a B-spline transform (final grid spacing 15\,mm, three
resolution levels), optimised with Mattes mutual information, a bending
energy penalty, and a rigidity penalty, using the same parameter
file\footnote{\url{https://github.com/SynthRAD2025/preprocessing}} as
the challenge preprocessing pipeline. These steps ensure that the MR and
CT images share the same voxel grid, so that tissue maps produced by
different segmentation methods can be compared voxelwise. The resulting
deformed CT was used as input for all subsequent CT segmentation steps. We note that this preprocessing was performed solely to enable the paired
comparison in this validation study; CTseg itself does not require any such
preprocessing when used in practice.

After preprocessing, of the 156 subjects from centres A and C, 97 were excluded during quality
review because they had incomplete brain coverage (the head-and-neck
radiotherapy field of view often truncates the vertex or skull base). The
remaining 59 subjects were processed by all the three below pipelines, where all three
methods succeeded for every subject, so no further exclusions were
necessary.

\subsection{Segmentation Pipelines}
\label{sec:segmentation}

Three segmentation pipelines were applied to each subject:

\begin{enumerate}
    \item \textbf{SPM on MRI (silver standard).} SPM12's unified
    segmentation~\citep{ashburner2005unified} was applied to each
    preprocessed MR image using default parameters, producing
    native-space probability maps for six tissue classes. We retained
    the GM, WM, and CSF maps as the silver-standard reference
    segmentations.

    \item \textbf{SPM on CT (baseline).} The same SPM12 pipeline was
    applied to each preprocessed CT image. Because SPM's tissue priors
    were designed for MRI, this represents an out-of-domain
    application. Bias field regularisation was increased
    (\texttt{biasreg\,=\,10}) to suppress bias field estimation, since
    CT images lack the intensity inhomogeneity characteristic of MRI.
    The forward deformation field was saved and used to produce a
    spatially normalised CT image via SPM's normalise-write module.

    \item \textbf{CTseg on CT.} CTseg was applied to each preprocessed
    CT image using the \texttt{mu\_CTseg\_spm15.nii} atlas, so that
    CTseg's deformation fields map directly into SPM space.
    Native-space tissue maps and modulated warped tissue maps were
    retained, and the estimated deformation field was used to produce
    a spatially normalised CT image.
\end{enumerate}

All three pipelines produced modulated warped GM, WM, and CSF tissue
maps in MNI space at 1.5\,mm resolution, used for the predictive
validation experiment.

\subsection{Evaluation}
\label{sec:metrics}

\subsubsection{Segmentation accuracy}
\label{sec:seg_accuracy}

For each tissue class (GM, WM, CSF), we compared the CT-based segmentations
against the MRI silver standard. Tissue probability maps were binarised at a
threshold of 0.5 before computing overlap and surface distance metrics. We
report three complementary measures:
\begin{itemize}
    \item \textbf{Dice coefficient}: volumetric
    overlap, ranging from 0 (none) to 1 (perfect).
    \item \textbf{95th~percentile Hausdorff distance
    (HD95)}: near-worst-case boundary
    error (mm), robust to outliers.
    \item \textbf{Average symmetric surface distance (ASSD)}: mean boundary
    displacement (mm).
\end{itemize}

\subsubsection{Spatial normalisation quality}
\label{sec:norm_quality}

Normalisation quality was assessed in three ways. First, the group-average
of all normalised CT images was computed separately for each deformation
(SPM-MR, SPM-CT, and CTseg); a sharper average indicates more consistent
normalisation. Second, we computed the voxelwise coefficient of variation
(CoV\,=\,$\sigma / |\mu|$) across subjects within a brain mask (mean
intensity $>$ 20\,HU); a lower mean CoV indicates more consistent
normalisation. Third, we directly compared the spatial accuracy of the
deformation fields by warping each subject's native-space MR tissue maps
(GM, WM, CSF) to atlas space using both the SPM-CT and CTseg
deformations, then computing Dice and ASSD against the SPM-MR
warped tissue maps (binarised at 0.5) as reference. This isolates the
quality of the deformation from differences in the segmentation model,
since all three methods warp the same native-space MR tissue maps.

\subsubsection{Brain volumetrics}
\label{sec:volumetrics}

TBV (GM\,+\,WM) and TIV (GM\,+\,WM\,+\,CSF) were computed for all three
methods by summing native-space tissue probability maps weighted by voxel
volume. Before summation, an intracranial mask was applied to exclude
voxels outside the brain. The mask was derived from the CTseg atlas by
computing the softmax tissue probabilities and thresholding the sum of
GM, WM, and CSF at 0.5. This atlas-space mask was then warped to each
subject's native space using the SPM-MR deformation field. The same mask
was applied to all three methods to ensure a fair comparison. This step is
necessary because head CT scans typically extend well below the skull
base, and without masking, tissue probability assigned to extracranial
structures (e.g.\ cervical spine white matter, facial soft tissue) can
inflate volume estimates, particularly for SPM, whose tissue priors were
designed for brain MRI and may assign non-negligible probabilities to
non-brain tissue when applied to CT. Agreement was assessed using ICC(3,1) (a two-way mixed-effects,
single-measures intraclass correlation assessing
consistency)~\citep{shrout1979intraclass}, Pearson
correlation, and Bland-Altman analysis~\citep{bland1986statistical}.

\subsubsection{Predictive validation of normalisation}
\label{sec:prediction}

As an implicit measure of normalisation quality, we evaluated how well
normalised tissue maps support sex classification, following the
paradigm of~\citep{brudfors2019superres}. The rationale is that better
spatial normalisation should yield more anatomically consistent feature
representations across subjects, improving discriminability in
downstream analyses. CTseg's diffeomorphic registration model is more
expressive than SPM12's default parameterisation, and its atlas was
learned jointly from MRI and CT data; if these richer deformations
produce more descriptive normalised tissue maps, this should be
reflected in predictive performance, potentially even outperforming
SPM applied to MRI. We focus on sex rather than age because sex is
discriminable largely from gross morphological size (total brain and
intracranial volume), which is preserved directly in the modulated
warped tissue maps, making it well-matched to a small-sample
($n = 58$) evaluation of normalisation quality. Age, by contrast, is
a continuous target whose signal lies in subtler distributed atrophy
patterns and would require a substantially larger cohort to recover
reliably from these features alone.

We used $n = 58$ subjects for the experiment (one of the 59 subjects
lacked sex metadata and was excluded),  comprising 44 males and 14 females (76\% male, age range 35–85 years, mean $63.8\pm11.1$). For each method, modulated
warped GM, WM, and CSF maps (i.e.\ tissue probability maps warped to
atlas space and scaled by the Jacobian determinant to preserve total
tissue volume) were smoothed (8\,mm FWHM), masked to intracranial
voxels, and concatenated into a single feature vector per subject.
These features were used directly (without dimensionality reduction)
in a kernel-based Gaussian process (GP) classification framework
implemented in PRoNTo~\citep{schrouff2013pronto}, which computes a
linear kernel from the feature vectors. Mean centering was performed
within each cross-validation fold to prevent information leakage. We
used 10-fold cross-validation with identical folds across methods.
Folds were assigned at random and were not stratified by sex; given
the cohort imbalance (44 male, 14 female), individual folds may
contain few or zero female subjects, contributing variance to the
per-fold accuracy estimate.


\subsubsection{Runtime}
\label{sec:runtime}

Processing time was recorded for each method and subject (segmentation
only, excluding I/O and warping).

\subsection{Statistical Analysis}
\label{sec:stats}

Paired Wilcoxon signed-rank tests (two-sided, Bonferroni-corrected for
three tissue classes) compared segmentation metrics between SPM-on-CT and
CTseg-on-CT; the same tests and correction procedure were applied to the
normalisation metrics (warped-tissue Dice and ASSD). Segmentation and
normalisation metrics are reported as median~[interquartile range]. Volumetric agreement is reported as ICC with
95\% confidence intervals, Pearson~$r$, and Bland-Altman bias with
95\%~limits of agreement. Significance level: $\alpha = 0.05$.

\section{Results}
\label{sec:results}

\subsection{Segmentation Accuracy}

\begin{figure}[!htbp]
    \centering
    \IfFileExists{\resultsdir/fig_segmentation_metrics.pdf}{%
        \includegraphics[width=\columnwidth]{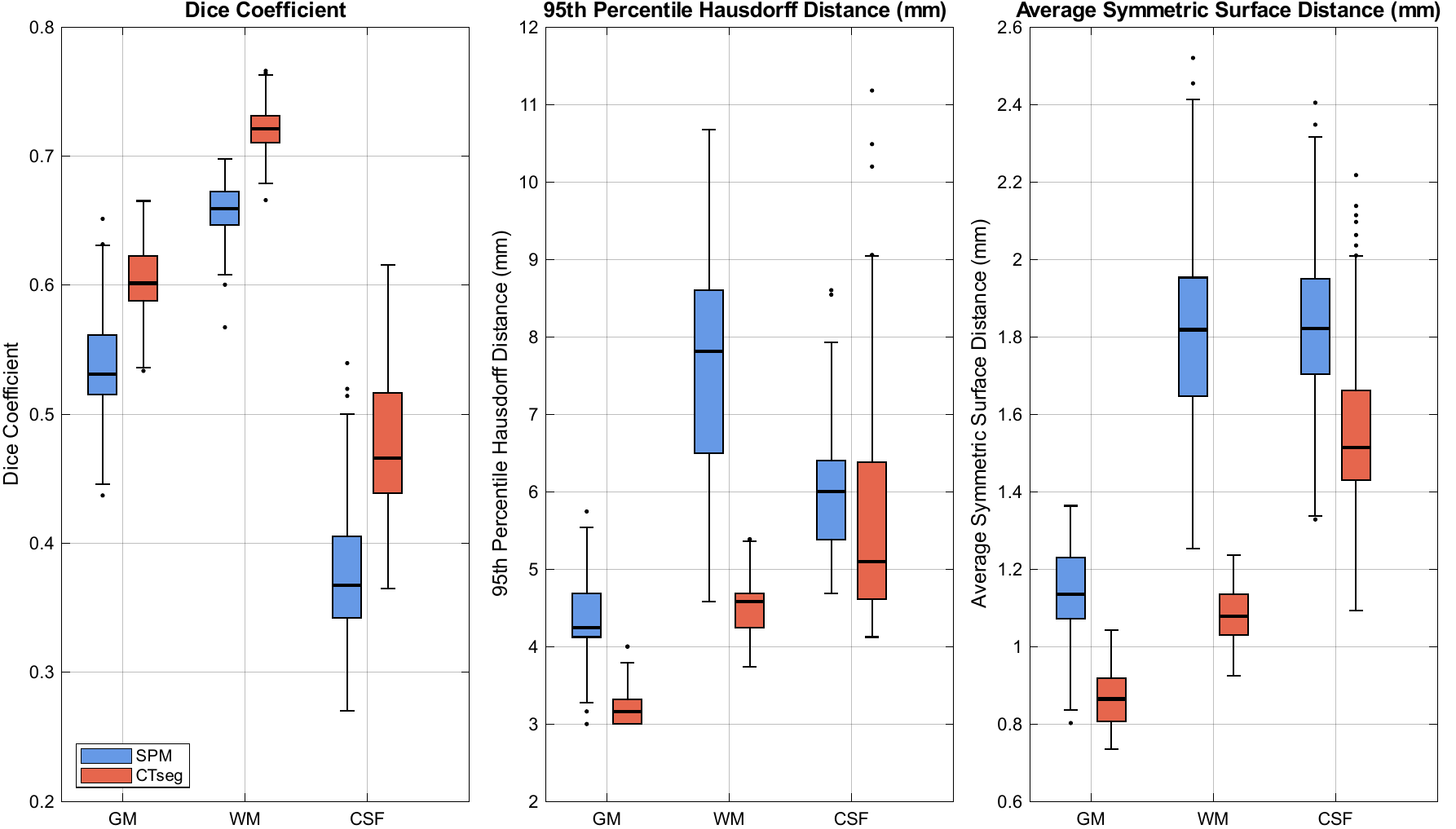}%
    }{\fbox{\parbox{\columnwidth}{\centering\vspace{2cm}fig\_segmentation\_metrics.pdf\vspace{2cm}}}}
    \caption{Segmentation metrics comparing SPM and CTseg on CT against the
    MRI silver standard. Left: Dice coefficient (higher is better). Centre:
    HD95 (lower is better). Right: ASSD (lower is better).}
    \label{fig:segmentation}
\end{figure}

\begin{figure}[!htbp]
    \centering
    \IfFileExists{\resultsdir/fig_example_tissue_maps.pdf}{%
        \includegraphics[width=0.6\columnwidth]{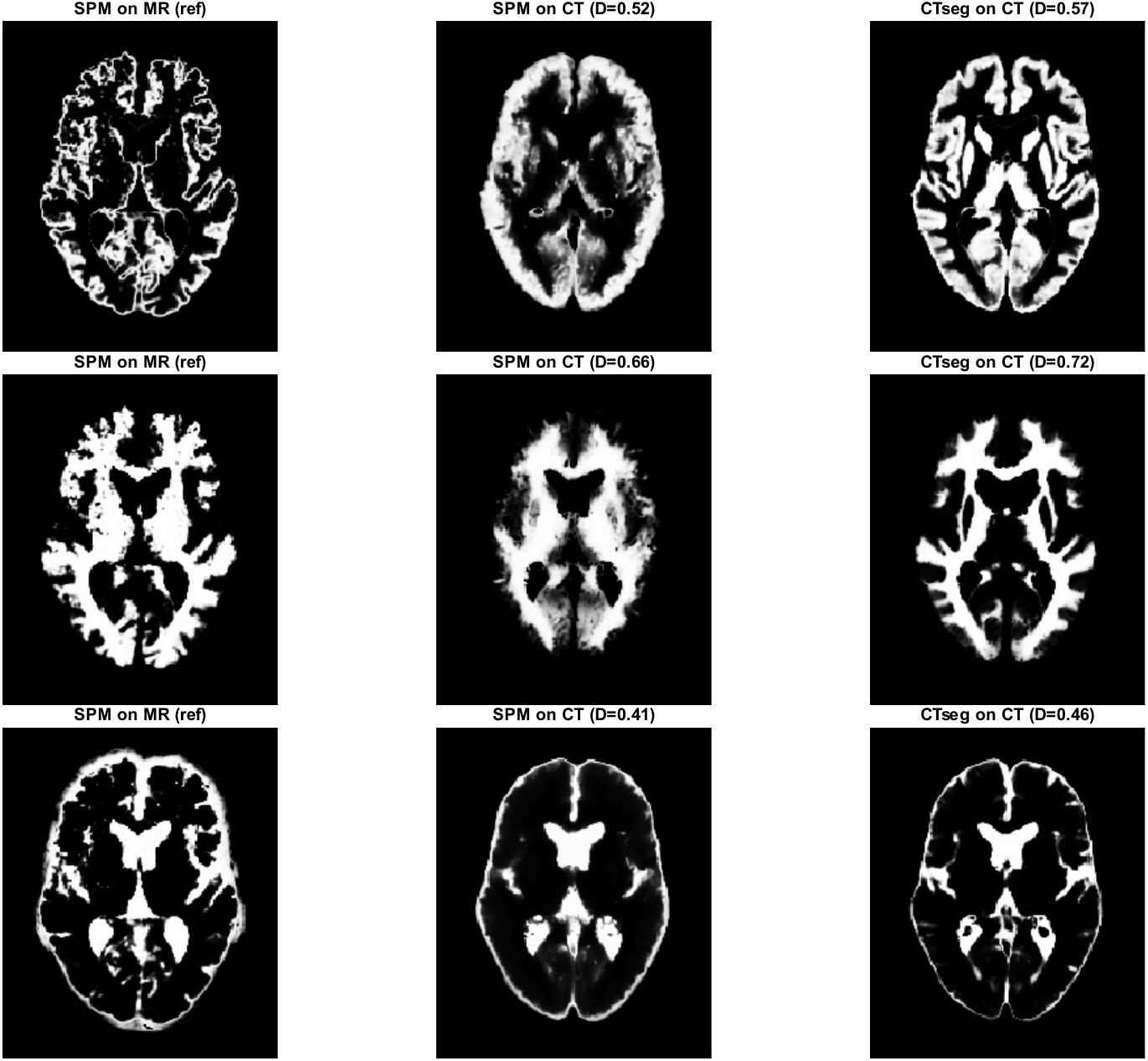}%
    }{\fbox{\parbox{\columnwidth}{\centering\vspace{2cm}fig\_example\_tissue\_maps.pdf\vspace{2cm}}}}
    \caption{Tissue probability maps for a representative subject (near-median
    Dice). Rows: GM, WM, CSF. Columns: SPM on MRI (silver standard), SPM on
    CT, CTseg on CT. Looking at, e.g., GM of CTseg, it clearly shows finer details in the cortex compared to SPM on CT.}
    \label{fig:tissue_maps}
\end{figure}

\begin{figure}[!htbp]
    \centering
    \IfFileExists{\resultsdir/fig_best_worst_segmentations.pdf}{%
        \includegraphics[width=0.8\columnwidth]{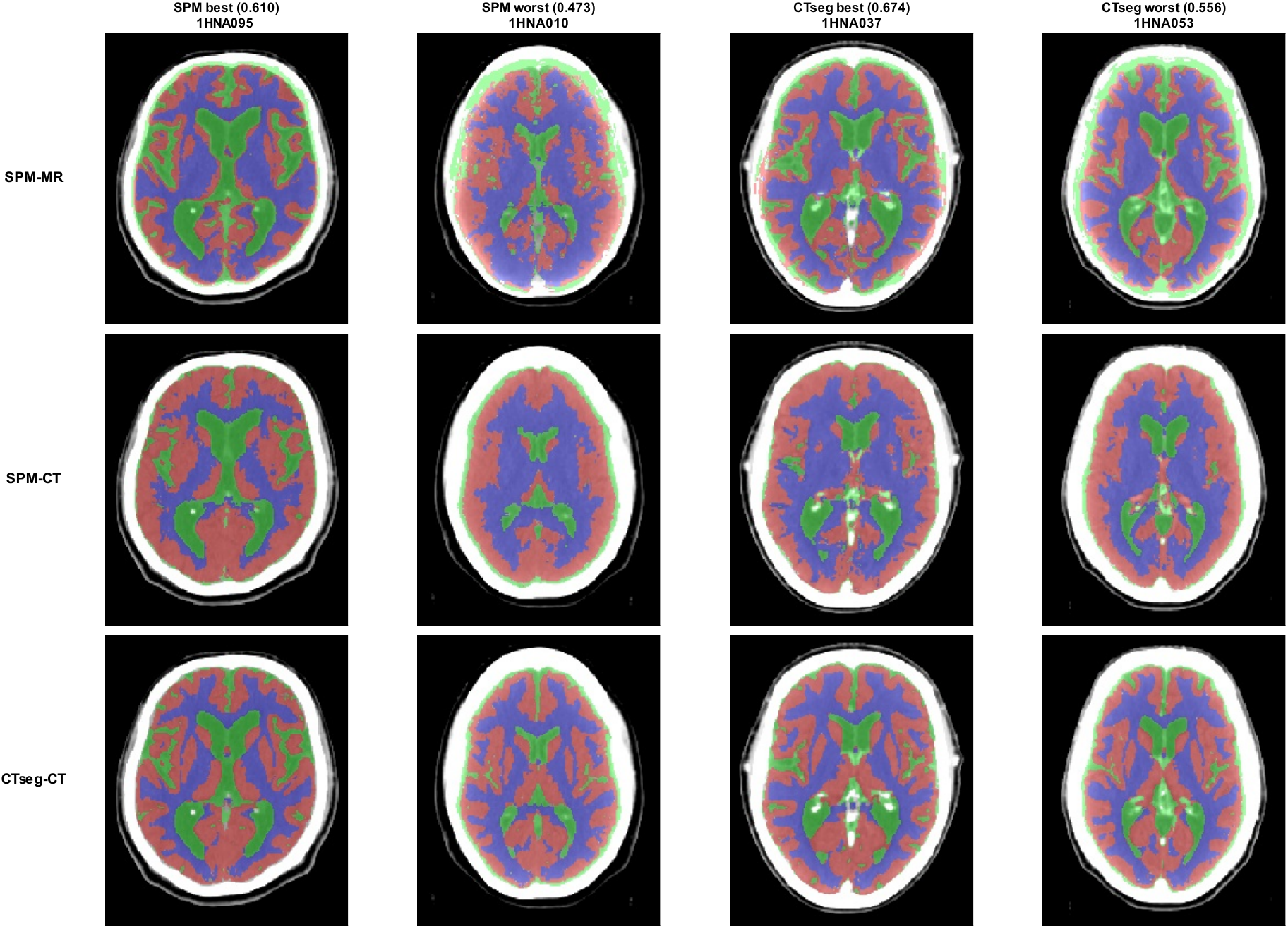}%
    }{\fbox{\parbox{\columnwidth}{\centering\vspace{2cm}fig\_best\_worst\_segmentations.pdf\vspace{2cm}}}}
    \caption{Segmentation overlays for best and worst cases (axial view).
    Rows: SPM on MRI (silver standard), SPM on CT, CTseg on CT. Columns:
    best and worst by SPM mean Dice (left pair), best and worst by CTseg
    mean Dice (right pair). Tissue colours: GM (red), WM (blue), CSF
    (green). Mean Dice scores are shown in column headers.}
    \label{fig:best_worst}
\end{figure}

Table~\ref{tab:segmentation} and Figure~\ref{fig:segmentation} summarise
the segmentation metrics across all subjects and tissue classes. CTseg
achieved significantly higher Dice scores and lower surface distances than
SPM-on-CT for all three tissue classes ($p < 0.001$ for all Dice and ASSD
comparisons; CSF HD95 did not reach significance).
Figure~\ref{fig:tissue_maps} shows tissue probability maps for a
representative subject, and Figure~\ref{fig:best_worst} shows segmentation
overlays for the best and worst cases.


\IfFileExists{\resultsdir/table_segmentation.tex}{%
    \begin{table}[t]
\centering
\caption{Segmentation accuracy compared to MRI silver standard. Values are median [IQR]. $p$-values from Wilcoxon signed-rank test (Bonferroni-corrected). Best value per column in bold.}
\label{tab:segmentation}
\small
\begin{tabular}{llccc}
\toprule
Metric & Method & GM & WM & CSF \\
\midrule
Dice & SPM & 0.531 [0.515, 0.561] & 0.659 [0.647, 0.672] & 0.367 [0.342, 0.405] \\
 & CTseg & \textbf{0.601 [0.588, 0.622]} & \textbf{0.721 [0.710, 0.731]} & \textbf{0.466 [0.439, 0.516]} \\
 & $p$-value & $<$0.001 & $<$0.001 & $<$0.001 \\
\midrule
HD95 (mm) & SPM & 4.243 [4.123, 4.690] & 7.810 [6.500, 8.602] & 6.000 [5.385, 6.403] \\
 & CTseg & \textbf{3.162 [3.000, 3.317]} & \textbf{4.583 [4.243, 4.690]} & \textbf{5.099 [4.610, 6.383]} \\
 & $p$-value & $<$0.001 & $<$0.001 & 0.099 \\
\midrule
ASSD (mm) & SPM & 1.135 [1.072, 1.229] & 1.818 [1.646, 1.953] & 1.821 [1.704, 1.949] \\
 & CTseg & \textbf{0.865 [0.807, 0.918]} & \textbf{1.078 [1.029, 1.136]} & \textbf{1.513 [1.430, 1.662]} \\
 & $p$-value & $<$0.001 & $<$0.001 & $<$0.001 \\
\bottomrule
\end{tabular}
\end{table}
}{\textcolor{red}{[Table: run experiments to generate table\_segmentation.tex]}}

\subsection{Spatial Normalisation}

\begin{figure}[!htbp]
    \centering
    \IfFileExists{\resultsdir/fig_average_normalised_box.png}{%
        \includegraphics[width=0.8\columnwidth]{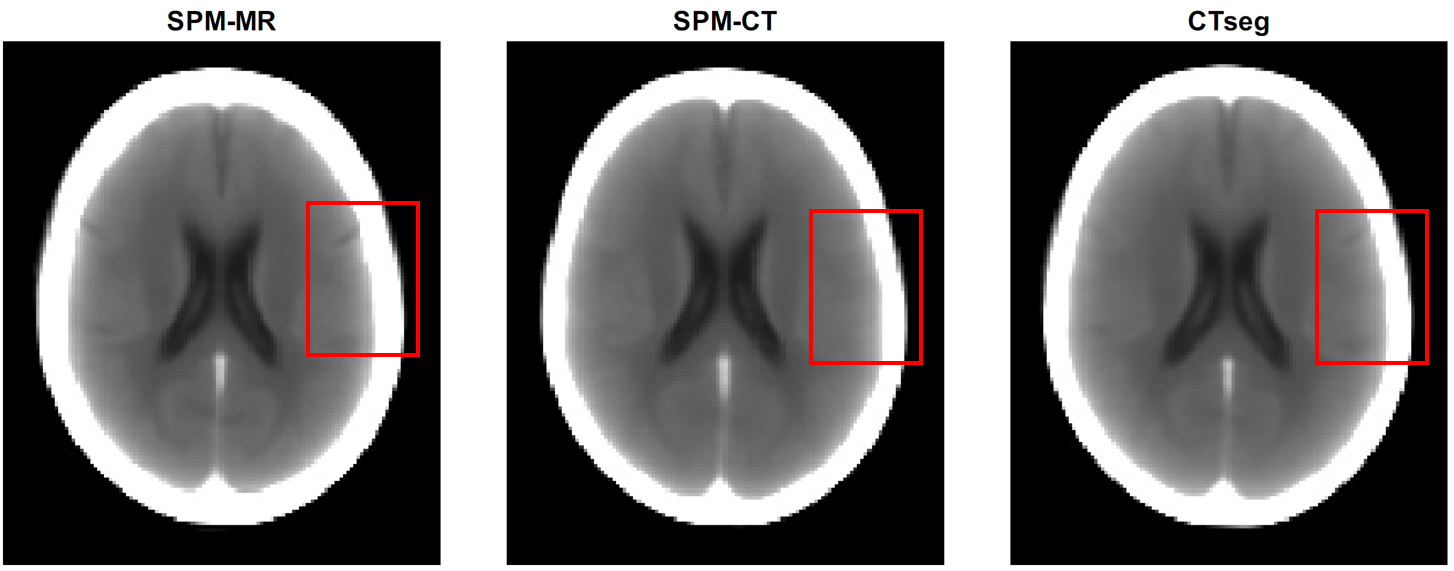}%
    }{\fbox{\parbox{\columnwidth}{\centering\vspace{2cm}fig\_average\_normalised.pdf\vspace{2cm}}}}
    \caption{Group-average normalised CT (axial view) for each deformation
    method. Sharper anatomical boundaries indicate more consistent spatial
    normalisation across subjects. Red boxes highlight the lateral
    ventricle and insular cortex region, where the CTseg average preserves
    cortical sulcal detail and the grey-white matter boundary comparably to
    the SPM-MR reference, whereas the SPM-CT average is visibly more
    blurred.}
    \label{fig:average}
\end{figure}

\begin{figure}[!htbp]
    \centering
    \IfFileExists{\resultsdir/fig_normalisation_metrics.pdf}{%
        \includegraphics[width=0.7\columnwidth]{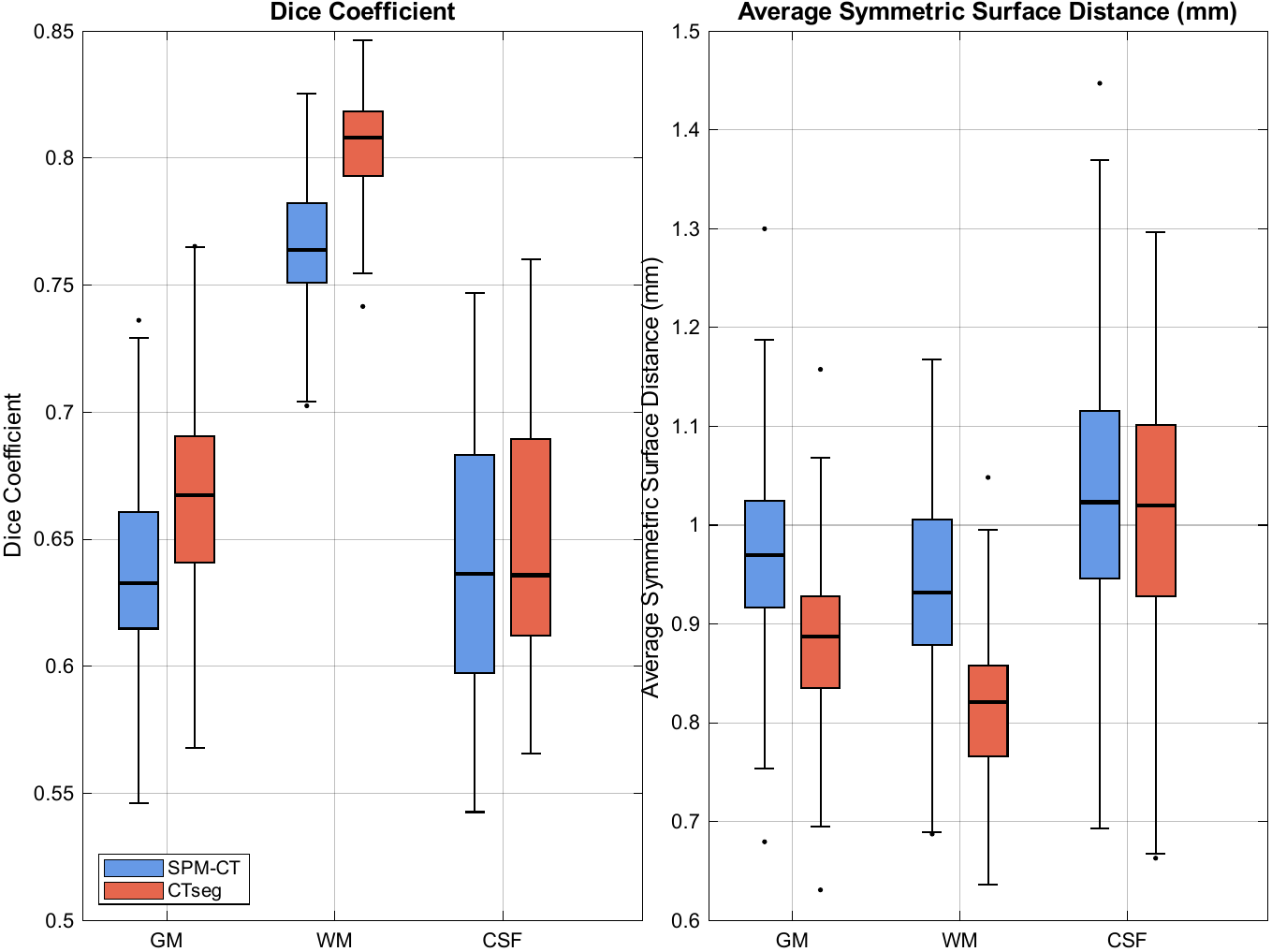}%
    }{\fbox{\parbox{\columnwidth}{\centering\vspace{2cm}fig\_normalisation\_metrics.pdf\vspace{2cm}}}}
    \caption{Normalisation metrics comparing SPM-CT and CTseg deformations.
    Native-space MR tissue maps were warped to atlas space using each
    deformation and compared against the SPM-MR warped tissue maps as
    reference. Left: Dice coefficient (higher is better). Right: ASSD
    (lower is better). HD95 is omitted because at the 1.5\,mm atlas
    resolution, the distance transform quantises surface distances into a
    few discrete values, making HD95 insensitive to differences between
    methods.}
    \label{fig:norm_metrics}
\end{figure}

Figure~\ref{fig:average} shows the group-average normalised CT images
computed using the three deformation fields. Visual inspection suggests
that the CTseg average preserves cortical detail and ventricular
boundaries more clearly than the SPM-CT average, consistent with more
consistent spatial normalisation across subjects. The mean voxelwise
coefficient of variation (CoV) within the brain mask was 0.306 for CTseg
versus 0.402 for SPM-CT, confirming more consistent alignment.
To quantify normalisation accuracy independently of segmentation
differences, we warped each subject's native-space MR tissue maps to
atlas space using both the SPM-CT and CTseg deformations, then
compared against the SPM-MR warped maps as reference
(Figure~\ref{fig:norm_metrics}). CTseg achieved significantly higher
Dice overlap than SPM-CT for GM (0.667 vs.\ 0.633, $p < 0.001$) and WM
(0.808 vs.\ 0.764, $p < 0.001$). For CSF, CTseg and SPM-CT gave
essentially identical median Dice (0.636 vs.\ 0.636, $p = 0.21$). ASSD
was significantly lower for CTseg in GM (0.887 vs.\ 0.970\,mm) and WM
(0.821 vs.\ 0.932\,mm), with no significant difference for CSF. These
results confirm that CTseg's deformation fields more accurately align
brain anatomy to atlas space than those obtained by applying SPM's
MRI-trained model to CT.

\subsection{Brain Volumetrics}

\begin{figure}[!htbp]
    \centering
    \IfFileExists{\resultsdir/fig_volumetrics_bland_altman.pdf}{%
        \includegraphics[width=0.8\columnwidth]{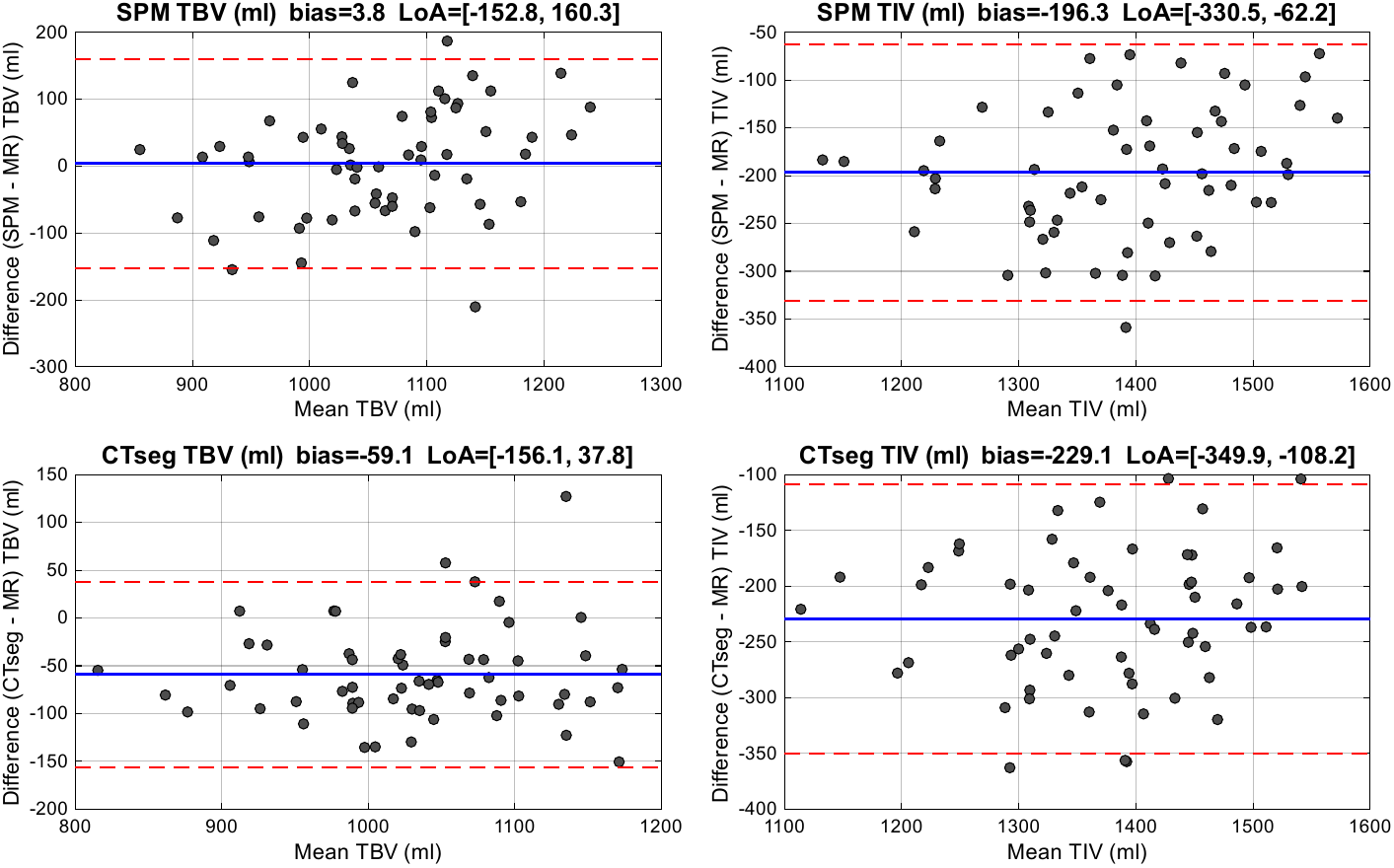}%
    }{\fbox{\parbox{\columnwidth}{\centering\vspace{2cm}fig\_volumetrics\_bland\_altman.pdf\vspace{2cm}}}}
    \caption{Bland-Altman plots for TBV (left) and TIV (right). Solid line:
    mean bias. Dashed lines: 95\% limits of agreement.}
    \label{fig:bland_altman}
\end{figure}

Table~\ref{tab:volumetrics} and Figure~\ref{fig:bland_altman} report
volumetric agreement between CTseg and MRI-derived estimates. For TBV,
CTseg achieves a substantially higher ICC (0.829 vs.\ 0.650) and tighter
limits of agreement than SPM-CT, indicating that it better preserves the
relative ordering of brain sizes across subjects. CTseg shows a systematic
negative bias of approximately $-59$\,ml, whereas SPM-CT's bias is near
zero ($3$\,ml). For TIV, both methods underestimate the MR-derived
reference by approximately 196-229\,ml, with similar ICC values (SPM-CT:
0.798, CTseg: 0.821). The TIV underestimation is expected given the
inherently poor contrast between CSF and brain parenchyma on CT.

If SPM-CT shows closer volumetric agreement with MRI-derived volumes
than CTseg does, this should be interpreted with caution: the MRI
reference volumes are themselves produced by SPM's unified segmentation,
which shares the same tissue probability atlas, Gaussian mixture
structure, regularisation, and cleanup pipeline as SPM-CT. Two runs of
the same generative model with the same priors will produce correlated
systematic biases, even when the input modality differs. Closer volume
agreement between SPM-CT and SPM-MR may therefore reflect shared-method
bias rather than genuinely superior accuracy. CTseg, by contrast, uses
a different atlas, different priors, and a different registration
framework (Multi-Brain), so its volume estimates are more independent
of the MRI-derived reference. The observed systematic bias suggests
that post hoc calibration may be feasible, but this was not evaluated
here; by contrast, the lower correlation of SPM-CT reflects poorer
rank-order agreement that cannot be corrected by a linear offset.
A further source of volume discrepancy specific to CTseg is its atlas
tissue class definitions: structures such as ventricular calcifications
(commonly found in the choroid plexus -- see CSF classes in Figure \ref{fig:tissue_maps}) and the venous sinuses are
modelled as non-brain classes (bone or soft tissue rather than CSF or
GM/WM), as visible in the CTseg atlas (see Figure~\ref{fig:template_original}).
Because these structures are intracranial but assigned to non-brain
tissue classes, they are excluded from TBV and TIV estimates,
contributing to the systematic negative bias relative to the MRI
reference.

\IfFileExists{\resultsdir/table_volumetrics.tex}{%
    \begin{table}[t]
\centering
\caption{Agreement of brain volume estimates (CT methods vs.\ MRI reference). Best value per column (within each measure) in bold; for Bias, smaller magnitude is better, for LoA, tighter interval is better.}
\label{tab:volumetrics}
\small
\begin{tabular}{llcccc}
\toprule
Measure & Method & ICC [95\% CI] & $r$ & Bias (ml) & LoA (ml) \\
\midrule
TBV & SPM & 0.650 [0.474, 0.776] & 0.670 & \textbf{3.8} & [-152.8, 160.3] \\
TBV & CTseg & \textbf{0.829 [0.729, 0.895]} & \textbf{0.830} & -59.1 & \textbf{[-156.1, 37.8]} \\
\midrule
TIV & SPM & 0.798 [0.683, 0.875] & 0.807 & \textbf{-196.3} & [-330.5, -62.2] \\
TIV & CTseg & \textbf{0.821 [0.716, 0.889]} & \textbf{0.823} & -229.1 & \textbf{[-349.9, -108.2]} \\
\bottomrule
\end{tabular}
\end{table}
}{\textcolor{red}{[Table: run experiments to generate table\_volumetrics.tex]}}

\subsection{Predictive Validation}

\begin{figure}[!htbp]
    \centering
    \IfFileExists{\resultsdir/fig_sex_roc.pdf}{%
        \includegraphics[width=0.5\columnwidth]{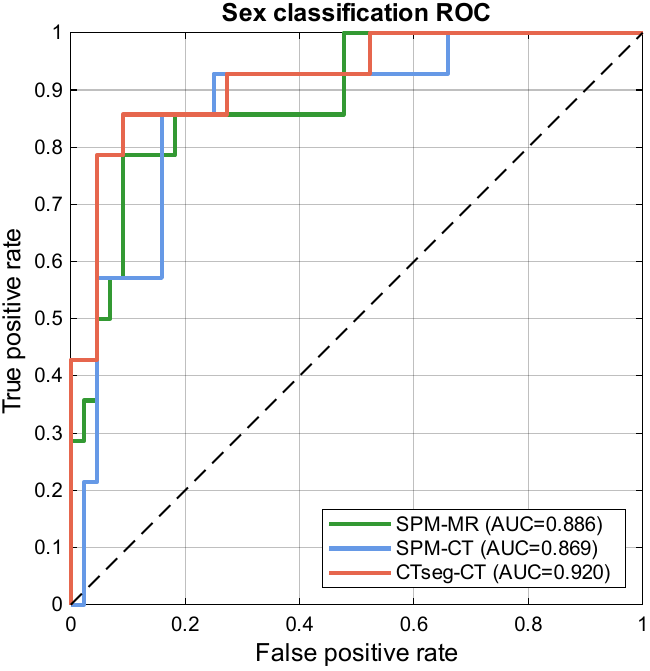}%
    }{\fbox{\parbox{0.6\columnwidth}{\centering\vspace{2cm}fig\_sex\_roc.pdf\vspace{2cm}}}}
    \caption{Sex classification ROC curves. Higher AUC indicates better
    normalisation quality.}
    \label{fig:sex_roc}
\end{figure}

For sex classification (Figure~\ref{fig:sex_roc}), CTseg achieved the
numerically highest AUC (0.920), with SPM-on-MR (0.886) and SPM-on-CT
(0.869) close behind, and the highest balanced accuracy (87.9\% vs.\
86.2\% and 84.5\%). AUC differences were not tested for statistical
significance given the small, sex-imbalanced cohort (44 male, 14
female; $n = 58$); the results should therefore be read as suggestive
rather than confirmatory. The pattern is nevertheless consistent with
the hypothesis that CTseg's more expressive diffeomorphic registration
preserves sex-discriminative morphological features (e.g.\ total brain
size, cortical thickness patterns) at least as faithfully as SPM's
parameterisation. AUC is threshold-independent and robust to the
imbalance: a classifier that simply predicts the majority class would
achieve 76\% accuracy but an AUC of only 0.5.

\subsection{Runtime}

All experiments were run on a workstation with an Intel Core Ultra~7 165H
(3.80\,GHz) processor and 32\,GB RAM, running Windows~11. No GPU
acceleration was used. Median segmentation time per subject was
84\,s (mean $83 \pm 11$\,s) for SPM-MR,
100\,s (mean $103 \pm 17$\,s) for SPM-CT, and
165\,s (mean $170 \pm 24$\,s) for CTseg ($n = 59$).
CTseg is approximately twice as slow as SPM, which is expected given its
more flexible nonlinear registration model (Multi-Brain diffeomorphic
framework versus SPM's DCT-based normalisation). All three methods
complete within minutes on a standard workstation without GPU
acceleration, making them practical for both research and clinical
deployment.

\section{Discussion}
\label{sec:discussion}

CTseg addresses a different problem from supervised CT labelling tools:
it provides a CT-specific generative pipeline for tissue segmentation,
spatial normalisation, and brain volumetrics within an SPM-compatible
framework, bridging the gap between the large volume of clinical CT data
and the quantitative analysis tools established for MRI. The goal of this
validation is to demonstrate that CTseg produces reliable outputs across
four complementary evaluation dimensions, substantially outperforming the
natural baseline of applying SPM's MRI-trained model to CT. Beyond this
formal validation, CTseg's practical utility is supported by its adoption
in over a dozen clinical studies spanning stroke, dementia, brain
morphometry, and other domains (Section~\ref{sec:adoption}).

For segmentation, CTseg achieved significantly higher Dice scores for all
three tissue classes (GM: 0.601 vs.\ 0.531; WM: 0.721 vs.\ 0.659; CSF:
0.466 vs.\ 0.367; all $p < 0.001$), with correspondingly lower surface
distances. These scores are lower than those reported for supervised deep
learning methods trained on paired CT/MRI
data~\citep{srikrishna2021deeplearning}, which is expected: CTseg is an
unsupervised generative model that requires no labelled training data and
jointly estimates segmentation, registration, and bias field parameters.
The trade-off is generality: CTseg is designed to work directly on
routine non-contrast CT scans, whereas supervised methods require
retraining for new populations, scanners, or acquisition protocols.
The largest absolute improvement over SPM-CT was in CSF, where the poor
CT contrast between CSF and parenchyma poses the greatest challenge for
MRI-trained models. That even GM and WM, which are better differentiated
on CT than CSF, show significant improvement underscores the value of
CT-specific intensity priors and atlas geometry.

Spatial normalisation quality showed a similar pattern: CTseg produced
sharper group averages and higher Dice overlap when the same native-space
MR tissue maps were warped using each method's deformation field. This
confirms that the improvement is not merely due to different segmentation
models but reflects genuinely more accurate anatomical alignment.

For brain volumetrics, CTseg achieved a substantially higher ICC for TBV
(0.829 vs.\ 0.650), indicating that it better preserves the relative
ordering of brain sizes across subjects. CTseg's systematic negative bias
($-59$\,ml for TBV, $-229$\,ml for TIV) is partially attributable to
its atlas class definitions and the inherently poor CSF contrast on CT
(discussed further in Limitations). The predictive validation provides a
complementary application-level assessment: CTseg achieved the
numerically highest sex classification AUC (0.920 vs.\ 0.869 for
SPM-CT and 0.886 for SPM-MR). Given $n = 58$ and the 76\%/24\% sex
imbalance this difference is suggestive rather than statistically
established, but the pattern is consistent with CTseg's more
expressive registration model producing normalised tissue maps that
preserve sex-discriminative morphological features at least as
faithfully as SPM applied to the higher-contrast MRI data.

\subsection{Limitations}
\label{sec:limitations}

\subsubsection{Study limitations}

Our MRI-derived reference segmentations constitute a \emph{silver}
standard, not a gold standard; both CT-based methods may be penalised
for disagreeing with the MRI reference where it is itself inaccurate.
We validate only GM, WM, and CSF, the classes for which MRI provides a
credible reference, and skull stripping, although one of CTseg's
outputs, was not directly evaluated. We also do not compare against
deep learning-based methods: as discussed in Section~\ref{sec:intro},
no established deep learning tool offers the same joint segmentation,
normalisation, and volumetric pipeline, and a comparison between
supervised and unsupervised paradigms would be better addressed in
dedicated future work.

The SynthRAD2025 dataset consists of head-and-neck radiotherapy
planning scans from cancer patients, not a general neuroimaging cohort.
Publicly available paired MR-CT brain datasets are exceedingly rare,
so co-registered same-subject data almost exclusively comes from
radiotherapy workflows; SynthRAD2025 is, to our knowledge, among the
largest such resources for the head region. Of the 156 available
subjects, 97 (62\%) were excluded for incomplete brain coverage (a
consequence of the radiotherapy planning field of view, which often
truncates the vertex or skull base), leaving 59 for quantitative
analysis. This limits statistical power and generalisability, although
the successful application of CTseg across diverse clinical
populations provides indirect evidence of broader applicability. The
predictive validation is limited by the small sample size ($n = 58$)
and the imbalanced sex distribution (76\% male) in this
retained cohort.

\subsubsection{Method limitations}

CT provides substantially lower contrast between CSF and brain
parenchyma than MRI, with the
CSF--cortex boundary being particularly ambiguous near cortical sulci
and the periventricular margin. CTseg's strong tissue prior partially
mitigates this, but CSF under-segmentation is an inherent ceiling of
CT-based brain segmentation rather than an algorithmic flaw. Since TIV
is defined as GM\,+\,WM\,+\,CSF, any CSF under-segmentation directly
reduces the TIV estimate, and TBV may also be affected if partial-volume
voxels at tissue boundaries are misclassified.

The CTseg atlas was learned from adult brain scans (CQ500, IXI,
MICCAI 2012, MRBrainS18). Paediatric and neonatal anatomy, and major
pathology not represented in these cohorts (large tumours, extensive
haemorrhage, post-surgical cavities, ventriculomegaly well beyond the
training distribution), are therefore outside the support of the
generative model. Under such distribution shift the unified-segmentation
framework can mis-assign voxels to the nearest-intensity tissue class,
producing plausible-looking but structurally incorrect segmentations.
Users working with populations far from the training distribution
should validate on representative cases before trusting CTseg outputs
quantitatively.

CTseg couples a Gaussian mixture on intensity with Multi-Brain's
diffeomorphic registration. This class of generative model has an
inherent accuracy ceiling relative to modern CNN/transformer
segmenters on in-distribution tasks with abundant labelled training
data. The trade-off is interpretability (each output is a posterior
under an explicit model), robustness to scanner/protocol variation
(no retraining required when moving between sites), data efficiency
(no labels), and a joint segmentation--registration--volumetrics
pipeline in a single call; CTseg is therefore best suited to settings
where generality and robustness matter more than pushing peak accuracy
on a fixed benchmark.

Finally, CTseg takes minutes per scan on a CPU, driven by iterative EM
and diffeomorphic registration, versus seconds for GPU-native CNN
segmenters. This is acceptable for research throughput and routine
offline clinical analysis, but CTseg is not intended for real-time
inference.

\subsection{Future Work}

Valuable extensions include multi-site validation, comparison with deep
learning approaches, validation of bone/soft-tissue classes, and evaluation
on pathological populations where robust CT segmentation is most needed.
Furthermore, CTseg's generative framework is not inherently specific to brain CT:
the tissue probability atlas and spatial normalisation machinery are
anatomy agnostic. By training Multi-Brain on, e.g., chest CT scans, the same
framework could be applied to different anatomies.

\bibliographystyle{plainnat}
\bibliography{references}

\end{document}